\documentclass[12pt,twoside]{article}

\usepackage{graphicx}
\usepackage{amssymb}

\catcode`@=11
\def\citer{\@ifnextchar
[{\@tempswatrue\@citexr}{\@tempswafalse\@citexr[]}}

\def\@citexr[#1]#2{\if@filesw\immediate\write\@auxout{\string\citation{#2}}\fi
  \def\@citea{}\@cite{\@for\@citeb:=#2\do
    {\@citea\def\@citea{--\penalty\@m}\@ifundefined
       {b@\@citeb}{{\bf ?}\@warning
       {Citation `\@citeb' on page \thepage \space undefined}}%
\hbox{\csname b@\@citeb\endcsname}}}{#1}}
\catcode`@=12

\newcommand{\st}{\tilde{t}}
\newcommand{\stb}{\bar{\tilde{t}}}
\newcommand{\sq}{\tilde{q}}
\newcommand{\sqb}{\bar{\tilde{q}}}
\newcommand{\gl}{\tilde{g}}
\newcommand{\gau}{\tilde{\chi}}
\newcommand{\MS}{\mbox{$\overline{\rm MS}$}}
\newcommand{\MSSM}{\mbox{$\MSSM}$}

\newcommand{\tgb}{\mbox{$\tan\beta$}}

 \newcommand{\zp}[3]{{Z.\ Phys.} {\bf #1} (19#2) #3}
 \newcommand{\np}[3]{{Nucl.\ Phys.} {\bf #1} (19#2)~#3}
 \newcommand{\pl}[3]{{Phys.\ Lett.} {\bf #1} (19#2) #3}
 \newcommand{\pr}[3]{{Phys.\ Rev.} {\bf #1} (19#2) #3}
 \newcommand{\prl}[3]{{Phys.\ Rev. Lett.} {\bf #1} (19#2) #3}

\begin{document}

\renewcommand{\thefootnote}{\fnsymbol{footnote}}
\setcounter{page}{0}

\begin{titlepage}

\begin{flushright}
DTP/98/88\\
CERN-TH/98-368\\
DESY 98-158\\
hep-ph/9810290 \\
October 1998
\end{flushright}
\vspace*{0.2cm}

\vspace*{0.8cm}

\begin{center} {\large \sc SUSY Particle Production at the
Tevatron\footnote{Contribution to the Workshop {\it Physics at Run II
-- Supersymmetry/Higgs}, 1998, Fermilab, USA.}}

\vspace*{1cm}

{\sc W.~Beenakker$^1$\footnote{Supported by a PPARC Research 
                               Fellowship}, 
 M.~Kr\"amer$^2$, 
 T.~Plehn$^3$\footnote{Supported in part by DOE grant 
                       DE-FG02-95ER-40896 and in part by the 
                       University of Wisconsin Research Committee 
                       with funds granted by the Wisconsin Alumni 
                       Research Foundation}, 
 and M.~Spira$^4$\footnote{Heisenberg Fellow}}

\vspace*{1cm}

$^1${\it Department of Physics, University of Durham, Durham DH1 3LE, U.K.}

$^2${\it Theory Division, CERN, CH-1211 Geneva 23, Switzerland}

$^3${\it Dept. of Physics, University of Wisconsin, Madison, USA}

$^4${\it II.\ Institut f\"ur Theoretische Physik\footnote{Supported by
Bundesministerium f\"ur Bildung und Forschung (BMBF), Bonn, Germany,
under Contract 05~7~HH~92P~(5), and by EU Program {\it Human Capital
and Mobility} through Network {\it Physics at High Energy Colliders}
under Contract CHRX--CT93--0357 (DG12 COMA).}, Universit\"at Hamburg,
Germany}

\end{center}

\vspace*{1cm}

\begin{abstract} The calculation of the next-to-leading order SUSY-QCD 
corrections to the production of squarks, gluinos and gauginos at the
Tevatron is reviewed. The NLO corrections stabilize the theoretical
predictions of the various production cross sections significantly and
lead to sizeable enhancements of the most relevant cross sections for
scales near the average mass of the produced massive particles. We
discuss the phenomenological consequences of the results on present
and future experimental analyses.
\end{abstract}

\end{titlepage}

\renewcommand{\thefootnote}{\arabic{footnote}}

\setcounter{footnote}{0}
\setcounter{page}{2}

\section{Introduction} 
%        ============ 

The search for supersymmetric particles is among the most important
endeavors of present and future high energy physics.  At the upgraded
$p\bar p$ collider Tevatron, the searches for squarks and gluinos, as
well as for the weakly interacting charginos and neutralinos, will
cover a wide range of the MSSM parameter space~\cite{CCRFM-97}.

%The novel colored particles, squarks and gluinos, and the weakly
%interacting gauginos can be searched for at the upgraded Tevatron, a
% with a c.m.\ energy of 2 TeV.  Until now the search
%at the Tevatron has set the most stringent bounds on the colored SUSY
%particle masses.  At the 95\% CL, gluinos have to be heavier than
%about 180 GeV, while squarks with masses below about 180 GeV have been
%excluded for gluino masses below $\sim 300$ GeV \cite{bounds}.  Stops,
%the scalar superpartners of the top quark, have been excluded in a
%significant part of the MSSM parameter space with mass less than about
%80 GeV by the LEP and Tevatron experiments \cite{bounds}.  Finally
%charginos with masses below about 90 GeV have been excluded by the LEP
%experiments, while the present search at the Tevatron is sensitive to
%chargino masses of about 60--80 GeV with a strong dependence on the
%specific model \cite{bounds}. Due to the negative search at LEP2 the
%lightest neutralino $\tilde \chi_1^0$ has to be heavier than about 30
%GeV in the context of SUGRA models \cite{bounds}. In the
%$R$-parity-conserving MSSM, supersymmetric particles can only be
%produced in pairs.  All supersymmetric particles will decay to the
%lightest supersymmetric particle (LSP), which is most likely to be a
%neutralino, stable thanks to conserved $R$-parity.  Thus the final
%signatures for the production of supersymmetric particles will mainly
%be jets, charged leptons and missing transverse energy, which is
%carried away by neutrinos and the invisible neutral LSP.

The cross sections for the production of SUSY particles in hadron
collisions have been calculated at the Born level already quite some
time ago \cite{LO}. Only recently have the theoretical predictions
been improved by calculations of the next-to-leading order SUSY-QCD
corrections \citer{sqgl,gaunlo}.  The higher-order corrections in
general increase the production cross section compared to the
predictions at the Born level and thereby improve experimental mass
bounds and exclusion limits. Moreover, by reducing the dependence of
the cross section on spurious parameters, {\it i.e.} the
renormalization and factorization scales, the cross sections in NLO
are under much better theoretical control than the leading-order
estimates.

The paper is organized as follows. In Section 2 we shall review the
calculation of the next-to-leading order SUSY-QCD corrections
\citer{sqgl,gaunlo}, by using the case of $\sq \sqb$ production as an
example. The NLO results for the production of squarks and gluinos are
presented in Section~3. We first focus on the scalar partners of the
five light quark flavors, which are assumed to be mass degenerate.
The discussion of final-state stop particles, with potentially large
mass splitting and mixing effects, is presented in Section 4.  In
Section 5 we discuss the NLO cross sections for the production of
charginos and neutralinos. We conclude the paper with a summary of the
relevant MSSM particle production cross sections at the upgraded
Tevatron, including next-to-leading order SUSY-QCD
corrections.\footnote{The MSSM Higgs sector will not be discussed
here, see instead Ref.\cite{MS-98}.}

\section{SUSY-QCD corrections}
%        ====================
The evaluation of the SUSY-QCD corrections consists of two pieces, the
virtual corrections, generated by virtual particle exchanges, and the
real corrections, which originate from real-gluon radiation as well as
from processes with an additional massless (anti)quark in the final
state.

\subsection{Virtual corrections}
%           ===================
The one-loop virtual corrections, {\em i.e.}\ the interference of the
Born matrix element with the one-loop amplitudes, are built up by
gluon, gluino, quark and squark exchange contributions (see
Fig.~\ref{fg:virt}a). We have adopted the fermion flow
prescription~\cite{fermion} for the calculation of matrix elements
including Majorana particles. The evaluation of the one-loop
contributions has been performed in dimensional regularization,
leading to the extraction of ultraviolet, infrared and collinear
singularities as poles in $\epsilon = (4-n)/2$. For the chiral
$\gamma_5$ coupling we have used the naive scheme, which is well
justified in the present analysis at the one-loop level.\footnote{We
have explicitly checked that the results obtained with a consistent
$\gamma_5$ scheme are identical to those obtained with the naive
scheme.}  After summing all virtual corrections no quadratic
divergences are left over, in accordance with the general property of
supersymmetric theories. The renormalization of the ultraviolet
divergences has been performed by identifying the squark and gluino
masses with their pole masses, and defining the strong coupling in the
$\overline{\rm MS}$ scheme including five light flavors in the
corresponding $\beta$ function. The massive particles, {\em i.e.}\
squarks, gluinos and top quarks, have been decoupled by subtracting
their contribution at vanishing momentum transfer \cite{decouple}. In
dimensional regularization, there is a mismatch between the gluonic
degrees of freedom (d.o.f. = $n-2$) and those of the gluino (d.o.f. =
$2$), so that SUSY is explicitly broken. In order to restore SUSY in
the physical observables in the massless limit, an additional finite
counter-term is required for the renormalization of the novel $\sq \gl
\bar q$ vertex. These counter-terms have been shown to render
dimensional regularization consistent with supersymmetry~\cite{count}.

\subsection{Real corrections} 
\begin{figure}[ht] \begin{center}
\includegraphics[angle=0,width=12cm]{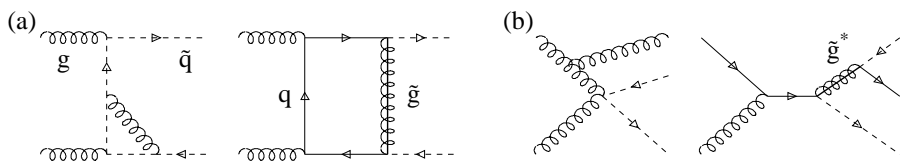}
\caption[]{\label{fg:virt} \label{fg:real} 
  \it Typical diagrams of the virtual (a) and real (b) corrections.}
\end{center} \end{figure}

%           ================ 
The real corrections are generated by real-gluon radiation off all
colored particles and by final states with an additional massless
(anti)quark, obtained from interchanging the final state gluon with a
light quark in the initial state (see Fig.~\ref{fg:virt}b).  The
phase-space integration of the final-state particles has been
performed in $n=4-2\epsilon$ dimensions, leading to the extraction of
infrared and collinear singularities as poles in $\epsilon$.  After
evaluating all angular integrals and adding the virtual and real
corrections, the infrared singularities cancel.  The left-over
collinear singularities are universal and are absorbed in the
renormalization of the parton densities at next-to-leading order. We
have defined the parton densities in the conventional $\overline{\rm
MS}$ scheme including five light flavors, {\em i.e.}\ the squark,
gluino and top quark contributions are not included in the mass
factorization. We finally obtain an ultraviolet, infrared and
collinear finite partonic cross section.

There is, however, an additional class of physical singularities,
which have to be regularized. In the second diagram of
Fig.~\ref{fg:real}b, an intermediate $\sq \gl^*$ state is produced,
before the (off-shell) gluino splits into a $q\sqb$ pair. If the
gluino mass is larger than the common squark mass, and the partonic
c.m.\ energy is larger than the sum of the squark and gluino masses,
the intermediate gluino can be produced on its mass-shell. Thus the
real corrections to $\sq \sqb$ production contain a contribution of
$\sq \gl$ production. The residue of this part corresponds to $\sq
\gl$ production with the subsequent gluino decay $\gl \to \sqb q$,
which is already contained in the leading order cross section of $\sq
\gl$ pair production, including all final-state cascade decays.  This
term has been subtracted in order to derive a well-defined production
cross section. Analogous subtractions emerge in all reactions: if the
gluino mass is larger than the squark mass, the contributions from
$\gl \to \sq \bar q, \sqb q$ have to be subtracted, and in the reverse
case the contributions of squark decays into gluinos have to
subtracted.\smallskip

\section{Results}
%        =======
In the following, we will present numerical results for SUSY particle
production cross sections at the upgraded Tevatron ($\sqrt{s}=2$~TeV),
including SUSY-QCD corrections.  The hadronic cross sections are
obtained from the partonic cross sections by convolution with the
corresponding parton densities. We have adopted the CTEQ4L/M parton
densities \cite{CTEQ} for the numerical results presented below. The
uncertainty due to different parametrizations of the parton densities
in NLO is less than $\sim 15$~\%. The average final state particle
mass is used as the central value of the renormalization and
factorization scales and the top quark mass is set to $m_t =
175$~GeV. The $K$-factor is defined as $K = \sigma_{NLO} /
\sigma_{LO}$, with all quantities ($\alpha_s(\mu_R)$, parton densities, 
parton cross section) calculated consistently in lowest and in
next-to-leading order.

\subsection{Production of Squarks and Gluinos}
\begin{figure}[ht] \begin{center}
\includegraphics[angle=270,width=12.5cm]{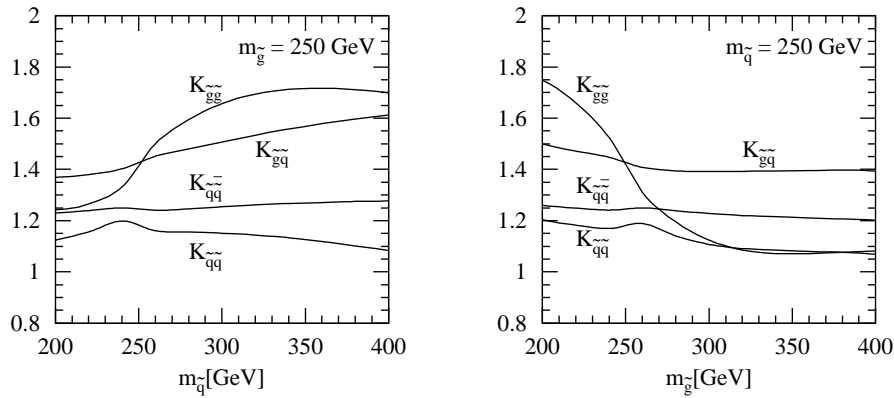}
\caption[]{\label{fg:kfac} 
  \it $K$-factors of the squark and gluino
      production cross sections at $Q=m$.}
\end{center} \end{figure}

%           =================================
Squarks and gluinos can be produced in different combinations via
$p\bar p \to \sq \sqb, \sq \sq, \sq \gl, \gl \gl$. We first focus on
the five light-flavored squarks, taken to be mass degenerate. At the
central renormalization and factorization scale $Q=m$, where $m$
denotes the average mass of the final-state squarks/gluinos, the
SUSY-QCD corrections are large and positive, increasing the total
cross sections in general by 10--90\% \cite{sqgl}. This is shown in
Fig.~\ref{fg:kfac}, where the $K$-factors are presented as a function
of the corresponding SUSY particle mass. The inclusion of SUSY-QCD
corrections leads to an increase of the lower bounds on the squark and
gluino masses by 10--30 GeV with respect to the leading-order
analysis.

\begin{figure}[ht] \begin{center}
\includegraphics[angle=270,width=12cm]{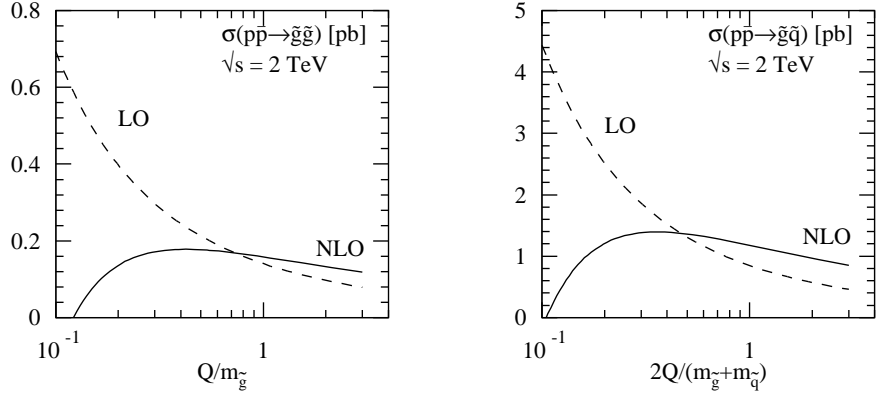}
\caption[]{\label{fg:scale} 
  \it Scale dependence of the total squark and gluino production 
      cross sections for 
      $m_{\sq}=250$~GeV and $m_{\gl}=300$~GeV.}
\end{center} \end{figure}

The residual renormalization/factorization scale dependence in leading
and next-to-leading order is presented in Fig.~\ref{fg:scale}. The
inclusion of the next-to-leading order corrections reduces the scale
dependence by a factor 3--4 relative to the lowest order and reaches a
typical level of $\sim 15\%$, when varying the scale from $Q = 2m$ to
$Q = m/2$. This may serve as an estimate of the remaining theoretical
uncertainty due to uncalculated higher-order terms.

\begin{figure}[ht] \begin{center}
\includegraphics[angle=270,width=6cm]{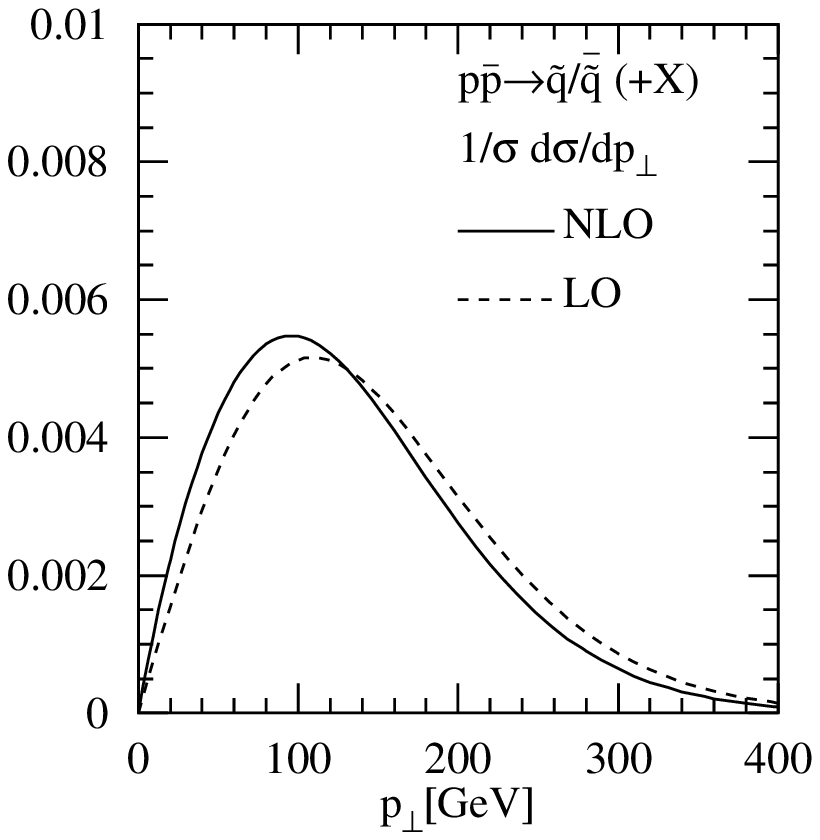} \hspace{0.5cm}
\includegraphics[angle=270,width=6cm]{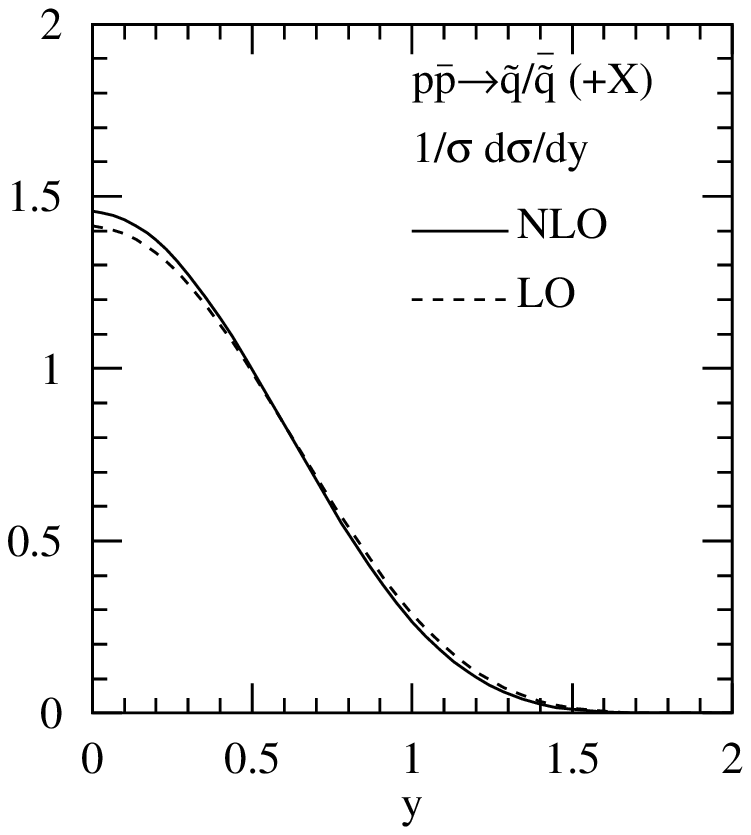}
\caption[]{\label{fg:pty} 
  \it Normalized transverse-momentum and rapidity
      distributions for squark production at $Q=m$.
      Mass parameters: $m_{\sq}=250$~GeV and $m_{\gl}=300$~GeV.}
\end{center} \end{figure}

Finally, we have evaluated the QCD-corrected single-particle exclusive
transverse-momentum and rapidity distributions for all different
processes.  As can be inferred from Fig.~\ref{fg:pty}, the
modification of the normalized distributions in next-to-leading
compared to leading order is less than about 15\% for the
transverse-mo\-men\-tum distributions and even smaller for the
rapidity distributions.  It is thus a sufficient approximation to
rescale the leading order distributions uniformly by the $K$-factors
of the total cross sections.

\subsection{Stop Pair Production} 
%           ==================== 

Stop production has to be considered separately since the strong
Yukawa coupling between top/stop and Higgs fields gives rise to
potentially large mixing effects and mass splitting. At leading-order
in the strong coupling constant $\alpha_s$, only diagonal pairs of
stop quarks can be produced in hadronic collisions, $p\bar p \to
\st_1\stb_1/\st_2\stb_2$. In contrast to the production of
light-flavor squarks, the leading-order $t$-channel gluino exchange
diagram is absent for stop production via $q\bar{q}$ initial states,
since top quarks are not included in the parton densities. The
leading-order stop cross section is thus in general significantly
smaller than the leading order cross section for producing
light-flavor squarks, where the threshold behavior is dominated by
$t$-channel gluino exchange. Mixed $\st_1 \st_2$ pair production can
safely be neglected since it can proceed only via one-loop $\alpha_s$
or tree level $G_F$ amplitudes and is suppressed by several orders of
magnitude \cite{stops}.  The evaluation of the QCD corrections
proceeds along the same lines as in the case of squarks and
gluinos.\footnote{The results obtained for the case of stop production
can also be used to predict the sbottom pair cross section at NLO
including mixing and mass splitting.} The strong coupling and the
parton densities have been defined in the \MS~scheme with five light
flavors contributing to their scale dependences, while the stop masses
are renormalized on-shell.

\begin{figure}[ht] \begin{center}
\includegraphics[angle=0,width=6cm]{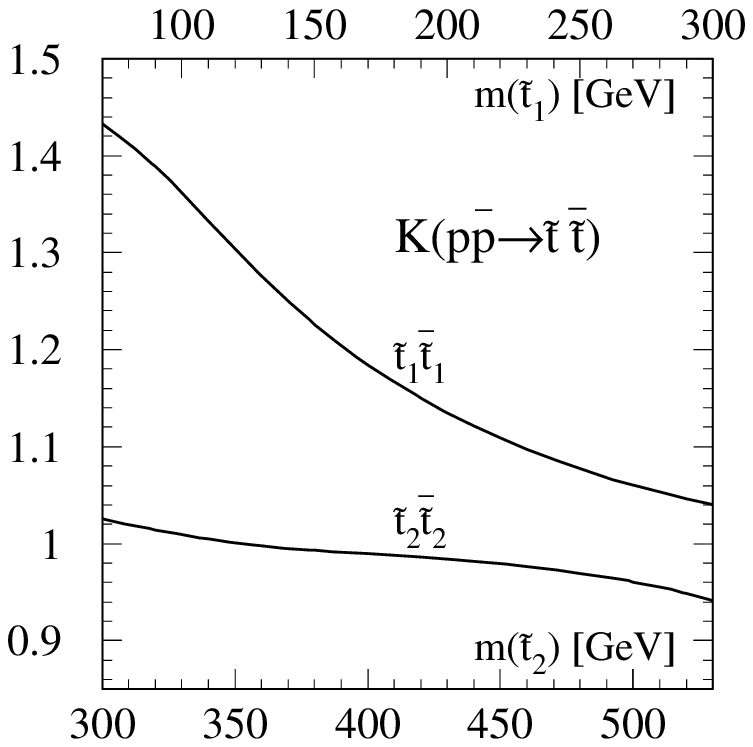} \hspace{0.5cm}
\includegraphics[angle=0,width=6cm]{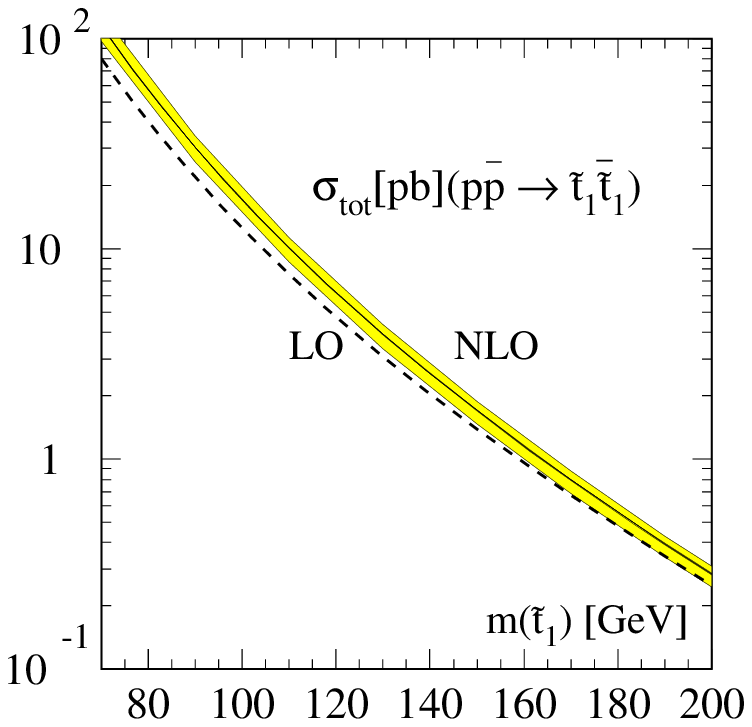}
\caption[]{\label{fg:kst} 
  \it Left: $K$-factor of the stop production cross sections at
      $Q=m_{\st}$ as a function of the stop masses [top/bottom scale].  
      Right: Production cross sections for the light stop
      state.  The thickness of the NLO curve represents the dependence
      of the cross sections on the stop mixing angle and the gluino
      and squark masses.  The shaded band indicates the theoretical
      uncertainty due to the scale dependence [$m/2<Q<2m$].}
\end{center} \end{figure}

The magnitude of the SUSY-QCD corrections is illustrated by the
$K$-factors at the central scale $Q=m_{\st}$ in Fig.~\ref{fg:kst}. In
the mass range relevant for the searches at the Tevatron, the SUSY-QCD
corrections are positive and reach a level of 30 to 45\% if the $gg$
initial state dominates. If, in contrast, the $q\bar{q}$ initial state
dominates, the corrections are small.  The relatively large mass
dependence of the $K$-factor for stop production at the Tevatron can
therefore be attributed to the fact that the $gg$ initial state is
important for small $m_{\tilde{t}}$, whereas the $q\bar{q}$ initial
state dominates for large $m_{\tilde{t}}$.

In complete analogy to the squark/gluino case, the scale dependence of
the stop cross section is strongly reduced, to about 15\% at
next-to-leading order in the interval $m_{\st}/2<Q<2m_{\st}$. The
virtual corrections at the NLO level depend on the stop mixing angle,
the squark and gluino masses, and on the mass of the second stop
particle. It turns out, however, that these dependences are very weak
for canonical SUSY masses and can safely be neglected, as can be
inferred from the light-stop production cross section in
Fig.~\ref{fg:kst}. On the other hand, internal particles with masses
smaller than the external particle mass, e.g. a light stop state
propagating in the loops for heavy stop production, will contribute to
the cross section.  This feature explains the small but noticeable
difference between the $\st_1$ and $\st_2$ $K$-factors at $m_{\st} =
300$~GeV shown in Fig.~\ref{fg:kst}.

\begin{figure}[ht] \begin{center}
\includegraphics[angle=0,width=6cm]{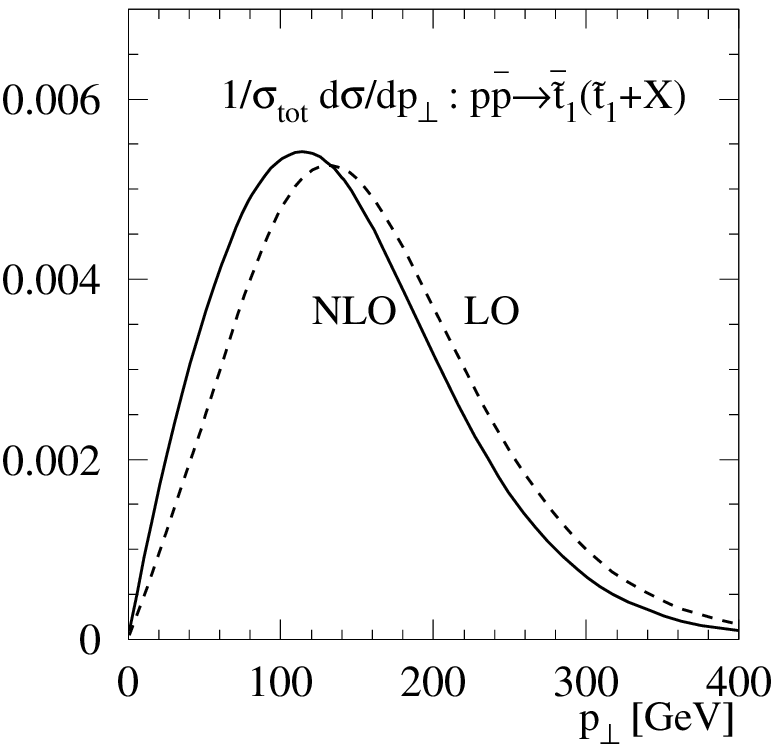} \hspace{0.5cm}
\includegraphics[angle=0,width=6cm]{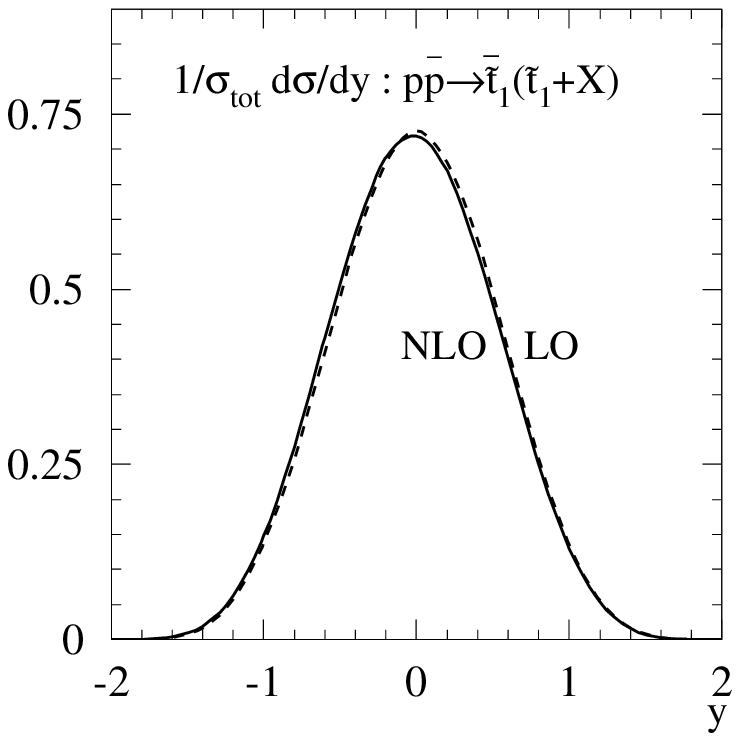}
\caption[]{\label{fg:tpty} 
  \it Normalized transverse-momentum and rapidity
      distributions for stop production at $Q=m_{\st}=200$~GeV.}
\end{center} \end{figure}

The next-to-leading order transverse-momentum and rapidity
distributions are presented in Fig.~\ref{fg:tpty}. While the shape of
the rapidity distribution is almost identical at leading and
next-to-leading order, the transverse momentum carried away by hard
gluon radiation in higher orders softens the NLO transverse momentum
distribution considerably.

\subsection{Chargino and Neutralino Production} 
%           ================================== 
At leading order, the production cross sections for chargino and
neutralino final states depend on several MSSM parameters, {\em i.e.}\
$M_1, M_2, \mu$ and $\tgb$ \cite{LO}. The cross sections are sizeable
for chargino/neutralino masses below about 100 GeV at the upgraded
Tevatron. Due to the strong sensitivity to the MSSM parameters, the
extracted bounds on the chargino and neutralino masses depend on the
specific region in the MSSM parameter space \cite{CCRFM-97}.  The
outline of the determination of the QCD corrections is analogous to
the previous cases of squarks, gluinos and stops. The resonance
contributions due to $gq \to \gau_i \sq$ with $\sq \to q \gau_j$ have
to be subtracted in order to avoid double counting of the associated
production of electroweak gauginos and strongly interacting squarks.
The parton densities have been defined with five light flavors
contributing to their scale evolution in the \MS~scheme, while the
$t$-channel squark masses have been renormalized on-shell.\footnote{
The next-to-leading order SUSY-QCD corrections to slepton pair
production can be trivially obtained from the corresponding results
for chargino/neutralino production. Numerically, the SUSY-QCD
corrections for slepton production agree with the pure QCD
corrections~\cite{sleptons}, provided the squark and gluino mass are
not chosen smaller than the final state slepton mass.}

\begin{figure}[ht] \begin{center} 
\includegraphics[angle=0,width=6cm]{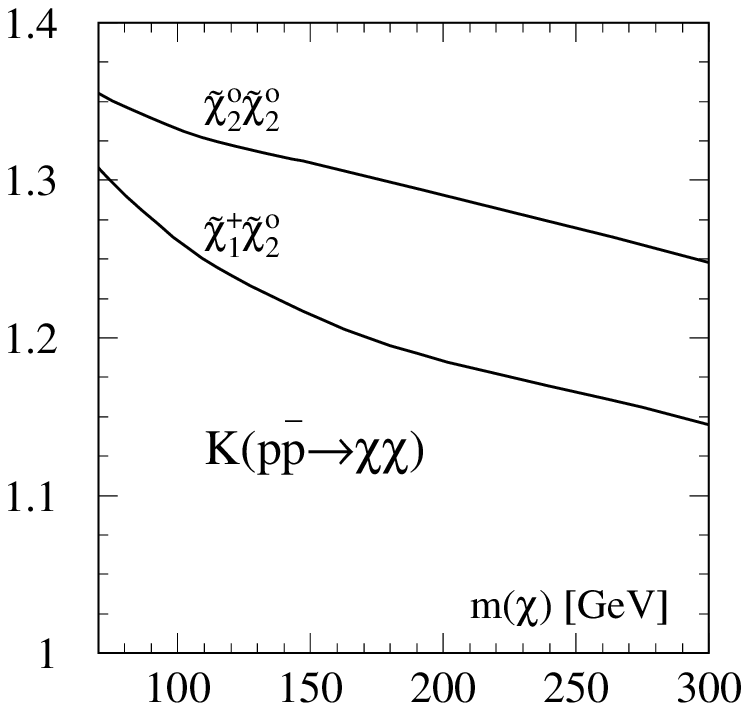} \hspace{0.5cm}
\includegraphics[angle=0,width=6cm]{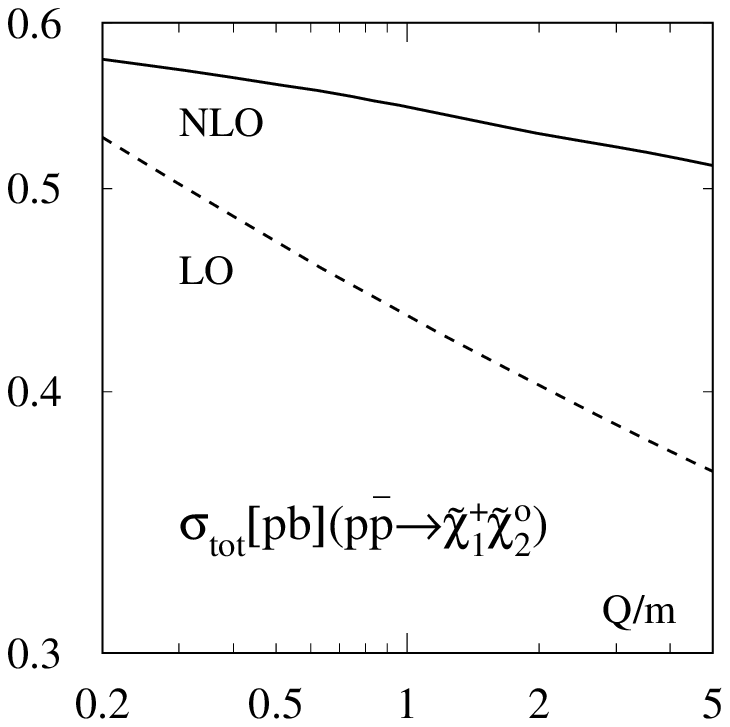}
\caption[]{\label{fg:kgausc} 
  \it Left: $K$-factor of the $\tilde{\chi}_2^{0}\tilde{\chi}_2^{0}$
  and $\tilde{\chi}_1^{+}\tilde{\chi}_2^{0}$ production cross sections
  at the central scale $Q=m$. Right: Scale dependence of the
  $\tilde{\chi}_1^{+}\tilde{\chi}_2^{0}$ cross section.  SUGRA
  parameters: $m_0 = 100$ GeV, $A_0 = 300$ GeV, $\tgb = 4$, $\mu >
  0$.}
\end{center} \end{figure}

At the average mass scale, the QCD corrections enhance the production
cross sections of charginos and neutralinos typically by about
10--35\% (see Fig.~\ref{fg:kgausc}), depending in detail on the final
state and the choice of MSSM parameters. The leading order scale
dependence is reduced to about 10\% at next-to-leading order (see
Fig.~\ref{fg:kgausc}), which implies a significant stabilization of
the theoretical prediction for the production cross sections
\cite{gaunlo}.

\begin{figure}[ht] \begin{center} 
\includegraphics[angle=0,width=6cm]{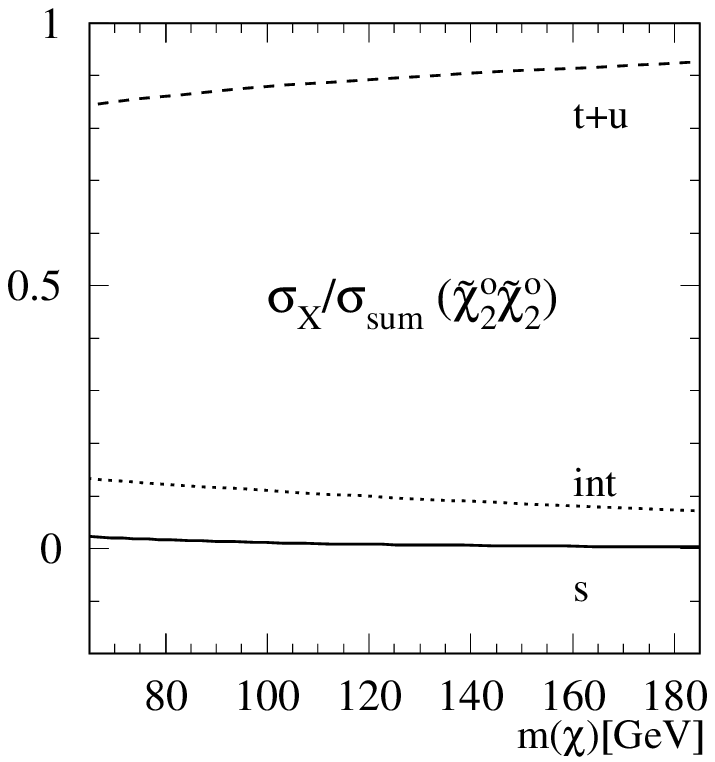} \hspace{0.5cm}
\includegraphics[angle=0,width=6cm]{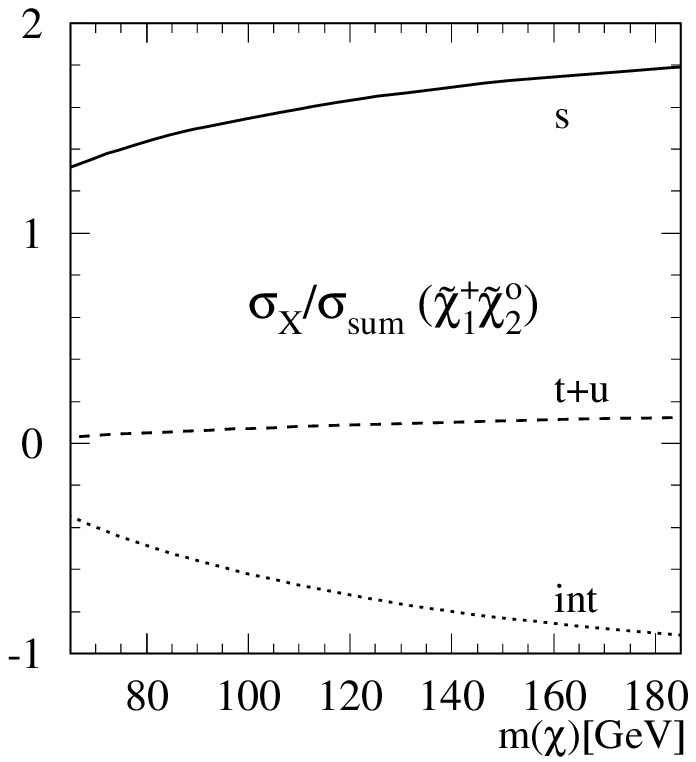}
\caption[]{\label{fg:gauind} 
  \it The $s$-channel, $(t+u)$-channel and interference contributions 
  to the 
  leading order $\tilde{\chi}_2^{0}\tilde{\chi}_2^{0}$ and 
  $\tilde{\chi}_1^{+}\tilde{\chi}_2^{0}$ production cross section.
  SUGRA parameters as in Fig.~\ref{fg:kgausc}}
\end{center} \end{figure}

The individual leading order contributions of the $s$-channel gauge
boson and the $t,u$-channel squark exchange are presented in
Fig.~\ref{fg:gauind}. For neutralino pair production the
$(t+u)$-channel contributions are by far dominating, while the
$s$-channel and interference terms are suppressed. Since the
$\gau^0_{1,2}$ states are predominantly gaugino-like, this reflects
the absence of a purely neutral trilinear gauge boson coupling in the
Standard Model. Contrary to that the $s$-channel of $\gau^\pm_1
\gau^0_2$ production is dominant and the $(t+u)$-channel term is
suppressed for large squark masses.  However, the interference turns
out to be sizeable.

\section{Conclusions} 
\begin{figure}[ht] \begin{center} 
\includegraphics[angle=0,width=11cm]{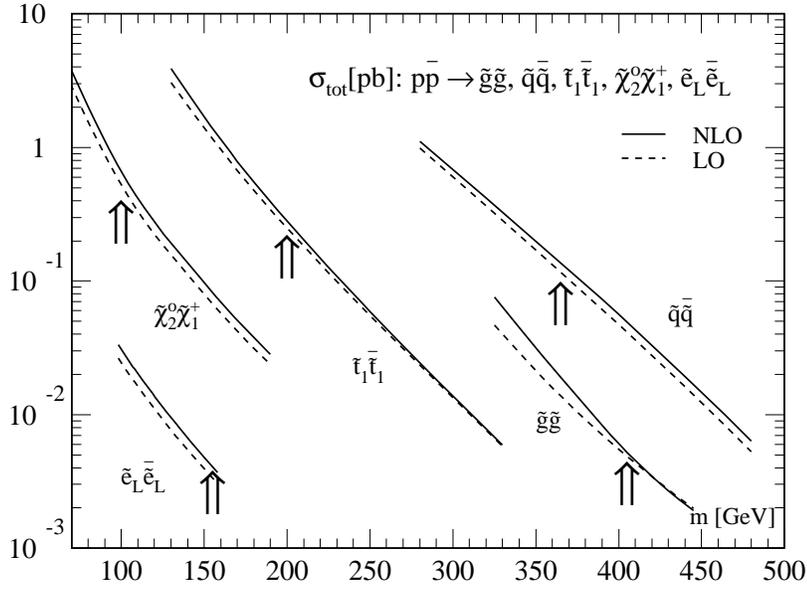}
\caption[]{\label{fg:sum_tev} 
  \it The NLO production cross sections included in {\tt PROSPINO} as
      a function of the final state particle mass; the arrows indicate
      the SUGRA inspired scenario: $m_{1/2}=150$~GeV, $m_0 = 100$~GeV,
      $A_0 = 300$~GeV, $\tgb = 4$, $\mu > 0$.  All cross sections are
      given at the average mass scale of the massive final-state
      particles.}
\end{center} \end{figure}

%        =========== 
We have reviewed the status of SUSY particle production at the
upgraded Tevatron at next-to-leading order in supersymmetric QCD. A
collection of relevant sparticle production cross sections is shown in
Fig.~\ref{fg:sum_tev}.  The higher-order corrections at the average
mass scale of the massive final-state particles significantly increase
the production cross section compared to the predictions at the Born
level. Experimental mass bounds are therefore shifted upwards.
Moreover, the theoretical uncertainties due to variation of
renormalization/factorization scales are strongly reduced to a level
of typically $\sim 15$~\%, so that the cross sections in
next-to-leading order SUSY-QCD are under much better theoretical
control than the leading order estimates. The NLO results for total
cross sections and differential distributions are available in the
form of the computer code {\tt PROSPINO} \cite{prospino}. \\

\noindent {\bf Acknowledgements} \\ We would like to thank P.M.\
Zerwas for his collaboration and encouragement. Moreover, we are
grateful to R.\ H\"opker and M.\ Klasen for their contribution to
different parts of this work.

\end{document}